\documentclass[astrosymb,twocolumn]{aastex631}

\usepackage{graphicx,xcolor}
\usepackage{tabularx}
\usepackage{dcolumn}
\usepackage{bm,bbm,xfrac,mathtools}
\usepackage[sans]{dsfont}
\usepackage{hyperref}
\usepackage[scr]{rsfso}
\usepackage{slashed}

\newcommand{\de}{\mathrm{d}}
\newcommand{\bs}{\ensuremath{b\sigma_8}}
\renewcommand{\fs}{\ensuremath{f\!\sigma_8}}
\newcommand{\lcdm}{\(\Lambda\)CDM}

\let\Gamma\varGamma
\let\Theta\varTheta
\let\Xi\varXi
\let\Pi\varPi
\let\Sigma\varSigma
\let\Upsilon\varUpsilon
\let\Phi\varPhi
\let\Psi\varPsi
\let\Omega\varOmega

\shortauthors{Tanidis \& Camera}

\begin{document}

\title{Model-independent constraints on clustering and growth of cosmic structures from BOSS DR12 galaxies in harmonic space}

\author{Konstantinos Tanidis}
\email{tanidis@fzu.cz}
\affiliation{CEICO, Institute of Physics of the Czech Academy of Sciences, Na Slovance 2, 18221 Praha 8, Czech Republic}
\affiliation{Dipartimento di Fisica, Universit\`a degli Studi di Torino, Via P.\ Giuria 1, 10125 Torino, Italy}
\affiliation{Istituto Nazionale di Fisica Nucleare -- INFN, Sezione di Torino, Via P.\ Giuria 1, 10125 Torino, Italy}

\author[0000-0003-3399-3574]{Stefano Camera}
\affiliation{Dipartimento di Fisica, Universit\`a degli Studi di Torino, Via P.\ Giuria 1, 10125 Torino, Italy}
\affiliation{Istituto Nazionale di Fisica Nucleare -- INFN, Sezione di Torino, Via P.\ Giuria 1, 10125 Torino, Italy}
\email{stefano.camera@unito.it}



\begin{abstract}
We present a new, model-independent measurement of the clustering amplitude of galaxies and the growth of cosmic large-scale structures from the Baryon Oscillation Spectroscopic Survey (BOSS) 12th data release (DR12). This is achieved by generalising harmonic-space power spectra for galaxy clustering to measure separately the magnitudes of the density and the redshift-space distortion terms, respectively related to the clustering amplitude of structures, \(\bs(z)\), and their growth, \(\fs(z)\). We adopt a tomographic approach with 15 redshift bins in \(z\in[0.15,0.67]\). We restrict our analysis to strictly linear scales, implementing a redshift-dependent maximum multipole for each bin. The measurements do not appear to suffer from systematic effects and show excellent agreement with the theoretical predictions from the Planck cosmic microwave background analysis assuming a \lcdm\ cosmology. Our results also agree with previous analyses by the BOSS collaboration. 
Furthermore, our method provides the community with a new tool for data analyses of the cosmic large-scale structure complementary to state-of-the-art approaches in configuration or Fourier space. Amongst its merits, we list: it being more agnostic with respect to the underlying cosmological model; its roots in a well-defined and gauge-invariant observable; the possibility to account naturally for wide-angle effects and even relativistic corrections on ultra-large scales; and the capability to perform an almost arbitrarily fine redshift binning with little computational effort. These aspects are all the more relevant for the oncoming generation of cosmological experiments such as Euclid, the Dark Energy Spectroscopic Instrument (DESI), the Legacy Survey of Space and Time (LSST), and the SKA Project.
\end{abstract}



\section{Introduction}
During the last couple of decades the physical picture of the cosmos has been set by the combination of the cosmic microwave background temperature and polarisation measurements \citep{Akrami2018} and large-scale structure probes. Among the most widely studied of these probes are the clustering of galaxies \citep{Beutler:2012px,Contreras_2013, Chuang2017, Mohammad_2018,descollaboration2021dark,prat2021dark,Porredon_2021,derose2021dark,porredon2021dark,pandey2021dark} and cosmic shear \citep{deJong2017,Aihara2019,jeffrey2021dark,secco2021dark,amon2021dark}.

All these data sets have led to a generally consistent picture summarised by the concordance \lcdm\ model, e.g.\ a Universe described by the late-time dark energy driving its accelerated expansion and by non-baryonic dark matter constituting the majority of the matter content today. Nonetheless, the nature of both dark energy and dark matter is far from understood. In addition, the appearance of tensions between different data sets \citep{Spergel2015,Addison2015,Battye2015,Raveri2016,Joudaki2016a,Joudaki2016b,PourtsidouTram2016,Charnock2017,Camera2017a} hints at possible cracks in the self-consistency of the \lcdm\ model.

One key tool to investigate the dark energy that has been widely used in the literature is redshift-space distortions (RSD). These trace the velocity field of matter inhomogeneities via the peculiar velocities of galaxies. They provide measurements on the growth rate of cosmic structures on large scales \citep{Kaiser:1987qv} and also affect the small (non-linear) scales due to incoherent galaxy motions within dark matter haloes. The former is known as the Kaiser effect, described by the squashing of the galaxy power spectrum perpendicular to the line-of-sight direction, whilst the latter phenomenon is dubbed the `Finger-of-God' effect, which enhances the power along the line-of-sight on small scales. The potential of RSD is excellent in discriminating between dark energy models since their growth rate prediction differs \citep[e.g.][]{2008Natur.451..541G}.

Typical analyses targeting RSD are usually done in Fourier- \citep{GilMarin2016,Zhao2017} or configuration-space \citep{Chuang2017,pellejeroibanez2016clustering,Wang2017}. These methods have to rely on assuming a fiducial cosmological model for the data, needed to transform redshifts to distances. Moreover, these kind of approaches are performed over broad redshift bins which can potentially hide valuable information on the expansion history \citep[but see e.g.][for a different approach]{2017MNRAS.464.2698R}.

Instead, in this paper we introduce a new method based on harmonic-space tomography in thin bins to infer the clustering and the growth rate as a function of redshift. Contrarily to the aforementioned approaches, the harmonic-space power spectrum is a natural observable for cosmological signals. We demonstrate that our method is robust against systematic effects and it also largely independent of the theoretical model.

As a case study, we choose to analyse the clustering of spectroscopic galaxies in the Baryon Oscillation Spectroscopic Survey (BOSS) 12th data release (DR12) \citep{Alam_2015}. In this context, it is worth mentioning \citet{Loureiro2019}, who used tomography in harmonic-space using BOSS data but focusing directly on the cosmological parameters estimation via model fitting, which is often referred to as full-shape analysis. On the other hand, we follow a template-fitting approach, since we are interested in comparing our novel harmonic-space approach to more standard analyses in Fourier space and configuration space. Note that a comparison of these two fitting methods is thoroughly discussed in e.g.\ \citet{Ivanov2020} and \citet{brieden2021shapefit}.

This paper is outlined as follows. In \autoref{sec:methodology}, we first present our new method and then discuss the data used in our analysis. In \autoref{sec:analysis}, we introduce the likelihood, covering first the construction of the theory and the data vectors, and later focusing on the data covariance matrix, which we estimate with three different approaches. The results are presented in \autoref{sec:result}, whilst in appendix~\ref{appendix} a series of consistency checks is performed. Finally, our concluding remarks are presented in \autoref{sec:conclusion}.



\section{Methodology}
\label{sec:methodology}
In this section we shall describe in detail the theoretical framework of our novel method and justify the specifics of our modelling. 
Then, we shall introduce the spectroscopic galaxies of BOSS DR12 \citep{Alam_2015} that we consider in our clustering analysis. That is, the specific data subsamples, the choice of their binning in redshift space and the construction of the final galaxy maps after taking into account various observational effects.

\subsection{Theory}
Up to linear order in cosmological perturbation theory, we can write the observed galaxy number count fluctuations in configuration space and in the longitudinal gauge as \citep{Yoo:2010ni,Challinor:2011bk,Bonvin:2011bg}
\begin{equation}
    \varDelta=b\,\delta-\frac{\partial_\parallel^2V}{\mathcal H}+\varDelta_{\rm loc}+\varDelta_{\rm int}\;.\label{eq:Delta_g}
\end{equation}
The first term represents matter density fluctuations, with \(\delta\) the density contrast in the longitudinal gauge, modulated by the linear galaxy bias \(b\). The second term is linear RSD, with \(\partial_\parallel\) the spatial derivative along the line-of-sight direction, \(\hat{\bm r}\), \(V\) the peculiar velocity potential, and $\mathcal H$ the conformal Hubble factor. The remaining two terms respectively collect all local and integrated contributions to the signal. Such terms are mostly subdominant and affect only the largest cosmic scales, reason for which we neglect them hereafter.\footnote{The weak gravitational lensing effect of cosmic magnification, factorised here in \(\varDelta_{\rm int}\), represents a notable exception to what just said. However, it matters mostly at high redshift and for large redshift bins and, since both such conditions are not met in the present analysis, we can safely ignore it.}

Since measurements from a galaxy survey corresponds to angular positions of galaxies on the sky, we can naturally decompose the observed \(\varDelta(\hat{\bm r})\) on the celestial sphere into its spherical harmonic coefficients, \(\varDelta_{\ell m}\). If additional information on the galaxies' redshift is available---as in the case of a spectroscopic galaxy survey---we can subdivide the data \(\varDelta(\hat{\bm r},z)\), which is now also a function of redshift, into concentrical redshift shells. This effectively corresponds to dealing with a set of \(\varDelta_{i,\ell m}=\int\de z\,n_i(z)\,\varDelta_{\ell m}(z)\), where the index \(i\) runs over the number of redshift bins and \(n_i(z)\,\de z\) is the number of sources per steradian in the \(i\)th redshift bin between \(z\) and \(z+\de z\). Note that \(n(z)\,\de z=n( r)\,\de r\) holds true, with \( r(z)\) the radial comoving distance to redshift \(z\).

Then, the harmonic-space tomographic power spectrum of galaxy number density fluctuations is
\begin{align}
    S^{\varDelta\varDelta}_{ij,\ell}\coloneqq&\left\langle\varDelta_{i,\ell m}\,\varDelta^\ast_{j,\ell m}\right\rangle\label{eq:Cl_DEF}\\
    =&\frac2\pi\,\int\de k\,k^2\,P_{\rm lin}(k)\,\varDelta_{i,\ell}(k)\,\varDelta_{j,\ell}(k)\;,\label{eq:Cl}
\end{align}
where \(k=|\bm k|\) is the Fourier mode corresponding to the physical separation between a pair of galaxies, \(P_{\rm lin}(k)\coloneqq P_{\rm lin}(k,z=0)\) is the present-day linear matter power spectrum, and
\begin{equation}
    \varDelta_{i,\ell}(k)=\int\de r\,n_i( r)\,D( r)\,\left[b( r)\,j_\ell(k r)-f( r)\,j^{\prime\prime}_\ell(k r)\right]\;.\label{eq:Delta_lk}
\end{equation}
In the equation above, \(D\) (normalised such that \(D=1\) at \(z=0\)) is the growth factor of density perturbations, \(f\coloneqq-\de\ln D/\de\ln(1+z)\) is the growth rate, \(j_\ell\) is the \(\ell\)th-order spherical Bessel function, and primes denote derivation with respect to the argument of the function.
%

Here, we are interested in developing a new analysis framework for the harmonic-space power spectrum of galaxy clustering, which can allow for a direct comparison to the results obtained with standard analyses of the three-dimensional Fourier-space power spectrum and real-space two-point correlation function. To do so, we adopt the common approach followed in these studies, in which constraints are obtained primarily on the bias, the growth, and the so-called distortion parameters related to the Alcock-Paczynski test \citep[see e.g.][\S~3.2.1]{2020A&A...642A.191E}. Note that, however, in a harmonic-space analysis we do not have the need to translate angles and redshifts into physical distances; we therefore are insensitive to the Alcock-Paczynski effect. Let us hence focus on expressing \autoref{eq:Cl} in terms of bias and growth.

First of all, it is useful to introduce the following two derived quantities:
\begin{align}
    \bs(z)&\coloneqq b(z)\,D(z)\,\sigma_8\;,\label{eq:bs8}\\
    \fs(z)&\coloneqq f(z)\,D(z)\,\sigma_8\;,\label{eq:fs8}
\end{align}
where \(\sigma_8^2\) is the rms variance of matter density fluctuations on spheres of \(8\,h^{-1}\,\mathrm{Mpc}\) radius; it effectively acts as a normalisation of \(P_{\rm lin}(k)\) at redshift zero. The definitions of \autoref{eq:bs8} and \autoref{eq:fs8} come from the consideration that, in a given redshift bin centred on \(\bar z\), the three quantities \(b\), \(f\), and \(\sigma_8\) are indistinguishable from each other in a measurement of the observed Fourier-space galaxy power spectrum
\begin{equation}
    P_{\varDelta\varDelta}(k,\mu;\bar z)\simeq\left[b(\bar z)+f(\bar z)\,\mu^2\right]^2\,D^2(\bar z)\,P_{\rm lin}(k)\;,
\end{equation}
where \(\mu=\hat{\bm r}\cdot\bm k/k\).

To recast \autoref{eq:Cl} in terms of \(\bs\) and \(\fs\), we require that these two quantities vary slowly across the redshift bin. Such a condition can be easily satisfied when dealing with spectroscopic redshift accuracy, which allows (in theory) for almost arbitrarily narrow redshift bins. Then, we factorise \(\bs\) and \(\fs\) out of the integral in \autoref{eq:Delta_lk} and thus write
\begin{equation}
    S^{\varDelta\varDelta}_{ij,\ell}\simeq\bs^i\,\bs^j\,T^{\delta\delta}_{ij,\ell}+\fs^i\,\fs^j\,T^{VV}_{ij,\ell}-2\,\bs^{(i}\,\fs^{j)}\,T^{\delta V}_{ij,\ell},\,\label{eq:Cl_ours}
\end{equation}
where, for a generic quantity \(X(z)\), we call \(X^i\) the value at the central redshift of the \(i\)th bin, round brackets around indexes denote symmetrisation, and we define the templates
\begin{align}
    T^{\delta\delta}_{ij,\ell}&\coloneqq\frac2\pi\,\int\de k\,\de r\,\de\tilde r\,k^2\,\frac{P_{\rm lin}(k)}{\sigma_8^2}\,n_i( r)\,n_j(\tilde r)\,j_\ell(k r)j_\ell(k\tilde r)\;,\label{eq:factorised_delta-delta}\\
    T^{VV}_{ij,\ell}&\coloneqq\frac2\pi\,\int\de k\,\de r\,\de\tilde r\,k^2\,\frac{P_{\rm lin}(k)}{\sigma_8^2}\,n_i( r)\,n_j(\tilde r)\,j^{\prime\prime}_\ell(k r)\,j^{\prime\prime}_\ell(k\tilde r)\;,\label{eq:factorised_v-v}\\
    T^{\delta V}_{ij,\ell}&\coloneqq\frac2\pi\,\int\de k\,\de r\,\de\tilde r\,k^2\,\frac{P_{\rm lin}(k)}{\sigma_8^2}\,n_{(i}( r)\,n_{j)}(\tilde r)\,j_\ell(k r)\,j^{\prime\prime}_\ell(k\tilde r)\;.
    \label{eq:factorised_delta-v}
\end{align}

%
Operatively, to compute the spectra defined above, we assume a Planck cosmology with parameters \citep{Ade2015}: \(\Omega_{\rm c}=0.2603\), \(\Omega_{\rm b}=0.0484\), \(H_0=67.74\,\mathrm{km\,s^{-1}\,Mpc^{-1}}\), \(\sigma_8=0.8301\), \(\tau=0.667\), and \(n_{\rm s}=0.9667\), them being, respectively, the cold dark matter abundance today, the baryon abundance today, the Hubble constant, the clustering amplitude, the optical depth to reionisation, and the slope of the primordial power spectrum of curvature perturbations.

\begin{figure}
    \centering
    \includegraphics[width=\columnwidth]{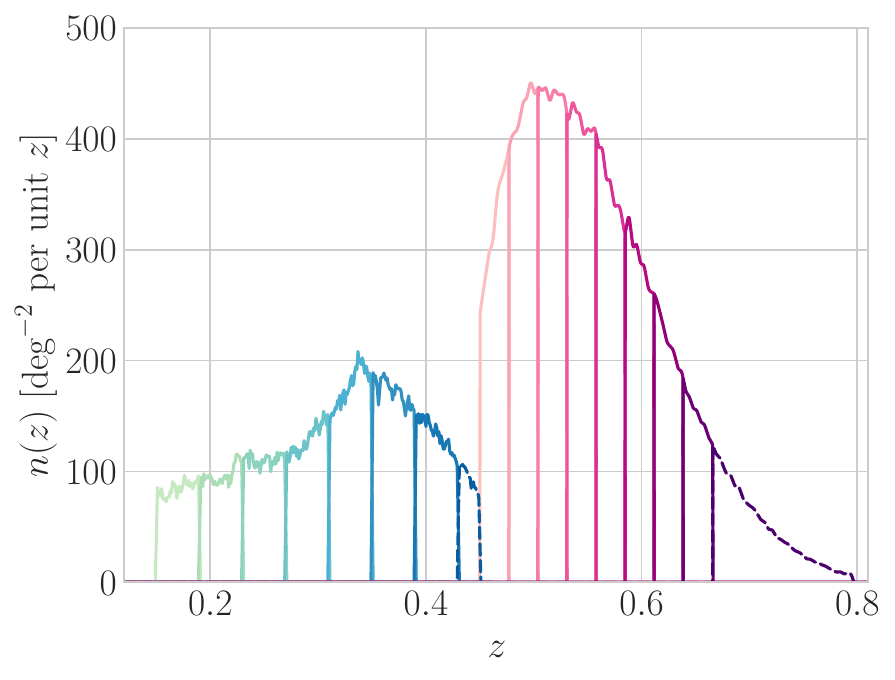}\\
    \includegraphics[width=\columnwidth]{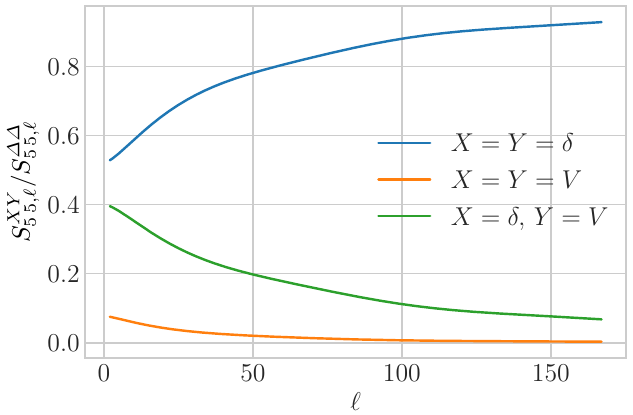}
    \caption{\textit{Top}: Redshift distributions and binning for the LOWZ (blue hues) and CMASS (pink hues) galaxy samples of BOSS DR12. Note that the highest-redshift bins of LOWZ and CMASS will not be included in the analysis (see \autoref{sec:data}) and are thus rendered here with dashed lines. \textit{Bottom}: Contributions to the total signal from the density (blue), RSD (yellow), and density-RSD (green) terms, for the \(i=j=5\) auto-bin spectrum of CMASS.}
    \label{fig:Nz_and_comparison_2}
\end{figure}

Moreover, we also need to specifcy the input redshift distribution of sources. For this, we consider both the LOWZ and CMASS spectroscopic galaxy samples of BOSS DR12 (for details see \autoref{sec:data}). Top panel of \autoref{fig:Nz_and_comparison_2} shows the two observed \(n(z)\)'s and the binning we adopt for LOWZ (blue hues) and CMASS (pink hues). We construct the top-hat bins as
\begin{equation}
n_i(z)= n(z) \times \frac12\,\left[1-\tanh\left({\frac{2\,|z-\bar{z}_i|-\varDelta z_i}{\sigma\,\varDelta z_i }}\right)\right],
    \label{eq:tophats}
\end{equation}
with \({\varDelta}z_i\) the bin width, \(\bar{z}_i\) the \(i\)th bin centre, and \(\sigma\) a smearing for the bin edges necessary to ensure numerical stability of the integration. For this, we choose \(\sigma=0.002\), and we have checked that the specific choice does not impact the results.

In the bottom panel of \autoref{fig:Nz_and_comparison_2} we present the three factorised spectra of \autoref{eq:factorised_delta-delta}-\ref{eq:factorised_delta-v}, modulated by their corresponding parameters as in the right-hand side of \autoref{eq:Cl_ours}, thus accounting for the density, RSD, and density-RSD contributions to the total signal (left-hand side of \autoref{eq:Cl_ours}). As a benchmark, we choose the CMASS auto-bin correlation with \(i=j=5\), and we plot the ratio between the factorised spectra and the total of \(S^{\varDelta\varDelta}_{ij,\ell}\). As expected, the main contribution comes from matter density fluctuations, with RSD being important only on very large scales. However, the cross-correlation between density and RSD is significantly non-negligible a term, especially at \(\ell\le50\) where it contributes to the total signal between \(20\%\) and \(40\%\).

\begin{figure}
    \centering
    \includegraphics[width=\columnwidth]{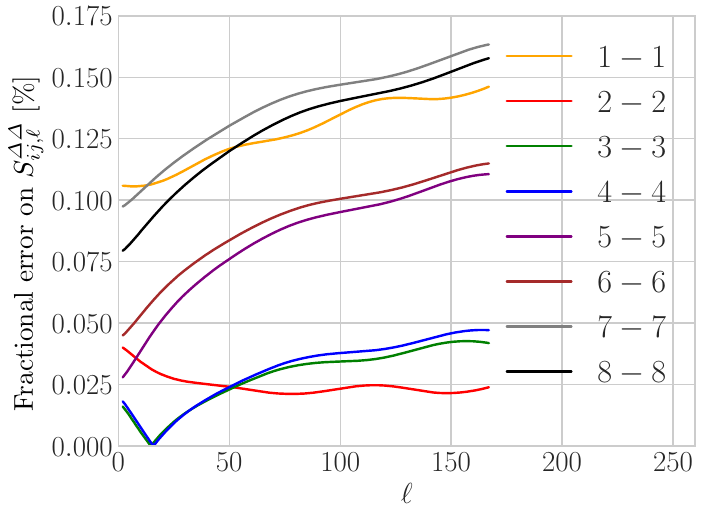}\\
    \includegraphics[width=\columnwidth]{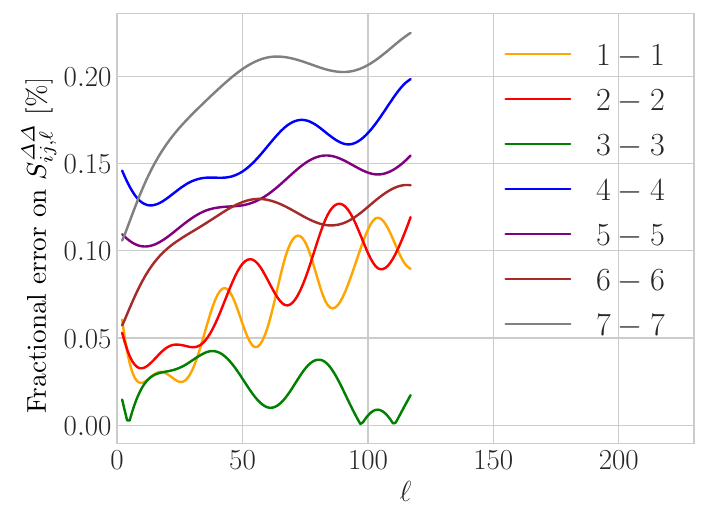}    \caption{Fractional error difference between the correct output of \autoref{eq:Cl} and our approximation of \autoref{eq:Cl_ours}, for LOWZ (top) and CMASS (bottom) auto-bin correlations.}
    \label{fig:comparison}
\end{figure}

To test the validity of our factorisation, in \autoref{fig:comparison} we demonstrate the performance of the approximated expansion described in \autoref{eq:Cl_ours} against the correct output of \autoref{eq:Cl}, for LOWZ in the top panel and CMASS in the bottom panel. (Redshift-bin ranges can be found in the second column of \autoref{tab:specs}.) The factorised spectra are calculated by modifying the publicly available code \texttt{CLASS} \citep{Lesgourgues2011}. It is evident that our approximation is in excellent agreement with the exact output at the subpercent level.

Finally, we emphasise that \autoref{eq:Cl} and \autoref{eq:Delta_lk} follow directly from \autoref{eq:Delta_g} and are therefore correct only at first-order in cosmological perturbation theory. Therefore, we restrict our analysis to strictly linear scales and fix the maximum wavenumber valid for linear theory to \(k_{\rm max}=0.1\,h\,\mathrm{Mpc}^{-1}\), independent of redshift. Then, we implement a conservative redshift-dependent maximum multipole \(\ell\) for each bin given by \(\ell_{\rm max}^i=k_{\rm max}\,r(\bar{z}_i)\). As for the minimum multipole, we define it by \(\ell_{\rm min}=\pi/(2f_{\rm sky})\), with \(f_{\rm sky}\) the survey sky coverage. In our specific case of \(f_{\rm sky}=0.2081\) for LOWZ and \(f_{\rm sky}=0.2416\) (see \autoref{sec:data}), this corresponds to a common \(\ell_{\rm min}=7\).

\subsection{Data}
\label{sec:data}
As already mentioned, in this study we use LOWZ and CMASS spectroscopic galaxies from the DR12 of BOSS. The LOWZ sample consists of Luminous Red Galaxies (LRGs) in the redshift range \(0.1<z<0.45\), whilst CMASS galaxies are LRGs at higher redshift, \(0.45<z<0.8\). Then, we have applied further redshift cuts on both samples, namely \(0.15\leq z \le 0.43\) for LOWZ and \(0.45\leq z \leq 0.67\) for CMASS. This choice is justified as follows. First, we ensure that the redshift ranges of the two samples do not overlap. Secondly, we make sure that the available mock catalogues from BOSS DR12 cover the same redshift range with our data selection. These mocks, as we will explain in detail in \autoref{subsec:Mock}, are essential for the investigation of systematic effects and for checking the pipeline's internal consistency. Finally, we avoid the inclusion of sources at \(z>0.67\) for CMASS so that we can ignore the small number of sources at that redshift range, which would introduce a considerable Poisson noise far surpassing the measured signal. 


Let us now describe the procedure followed to construct the masks and the galaxy overdensity maps for LOWZ and CMASS. All details are summarised in \autoref{tab:specs}.
\begin{table}
\centering
\caption{Redshift range, number of sources, shot noise, and maximum multipole for the redshift bins considered in the analysis.}
\setlength{\tabcolsep}{4.5pt}
\renewcommand{\arraystretch}{1.1}
 \begin{tabularx}{\columnwidth}{l l l l X}
 \hline
 Bin ID &  \([z_{\rm min},\, z_{\rm max}]\) & \# of gals  & shot noise \(\mathrm{[sr]}\) & \(\ell_{\rm max}^i\) \\  
\hline
 LOWZ-1 & \([0.150,\, 0.190]\) & \(33\,906\) & \(8.08\times10^{-5}\) & \(49\) \\
 LOWZ-2 & \([0.190,\, 0.230]\) & \(38\,728\) & \(7.07\times10^{-5}\) & \(60\) \\ 
 LOWZ-3 & \([0.230,\, 0.270]\) & \(44\,291\) & \(6.18\times10^{-5}\) & \(71\) \\
 LOWZ-4 & \([0.270,\, 0.310]\) & \(51\,781\) & \(5.27\times10^{-5}\) & \(81\) \\
 LOWZ-5 & \([0.310,\, 0.350]\) & \(70\,879\) & \(3.86\times10^{-5}\) & \(91\) \\
 LOWZ-6 & \([0.350,\, 0.390]\) & \(68\,701\) & \(3.98\times10^{-5}\) & \(101\) \\
 LOWZ-7 & \([0.390,\, 0.430]\) & \(53\,191\) & \(5.15\times10^{-5}\) & \(111\) \\
 CMASS-1 & \([0.450,\, 0.477]\) & \(81\,836\) & \(3.71\times10^{-5}\) & \(123\) \\ 
 CMASS-2 & \([0.477,\, 0.504]\) & \(112\,589\) & \(2.67\times10^{-5}\) & \(129\) \\ 
 CMASS-3 & \([0.504,\, 0.531]\) & \(118\,915\) & \(2.55\times10^{-5}\) & \(135\) \\
 CMASS-4 & \([0.531,\, 0.558]\) & \(113\,139\) & \(2.68\times10^{-5}\) & \(141\) \\
 CMASS-5 & \([0.558,\, 0.585]\) & \(99\,672\) & \(3.04\times10^{-5}\) & \(147\) \\
 CMASS-6 & \([0.585,\, 0.612]\) & \(80\,914\) & \(3.75\times10^{-5}\) & \(153\) \\
 CMASS-7 & \([0.612,\, 0.639]\) & \(61\,551\) & \(4.93\times10^{-5}\) & \(159\) \\
 CMASS-8 & \([0.639,\, 0.670]\) & \(44\,057\) & \(6.89\times10^{-5}\) & \(164\) \\
 \hline
\end{tabularx}
\label{tab:specs}
\end{table}

Besides the galaxy catalogues, the BOSS collaboration has made publicly available random catalogues that are fifty times denser than the galaxy ones. Those were constructed after taking into account the completeness and the veto masks. The completeness masks indicate to which extent the observations under consideration are complete, whilst the veto masks account for observational effects such as bright objects and stars, extinction and seeing cuts, fiber collisions, fiber centerposts, and others. We built the final binary mask based on the random catalogues, assigning 0 where there are no observations and 1 otherwise using the \texttt{HEALPix} pixelisation scheme \citep{Gorski_2005} with \(N_{\rm side}=1024\). We deem this resolution more than sufficent for the scales of interest in this work (last column of \autoref{tab:specs}), since the largest scale that can be safely resolved by a \texttt{HEALPix} map is \(\ell_{\rm max}=2N_{\rm side}\) \citep{Ando2017}.

Then, we proceed with the the construction of the galaxy overdensity maps (where the mask has value 1) for the 7 redshift bins of LOWZ and the 8 bins of CMASS as
\begin{equation}
    \varDelta_p = \frac{n_{{\rm g},p}-\langle n_{\rm g}\rangle_p}{\langle n_{\rm g}\rangle_p}\;,
    \label{eq:overdensity}
\end{equation}
where \(n_{{\rm g},p}\) is the number of weighted galaxies in a given pixel \(p\) and \(\langle n_{\rm g}\rangle_p\) is the mean number of weighted galaxies per pixel. We should note that the BOSS systematic weights \citep{Reid2016} are not included in our analysis since we checked that they do not impact our final results.

\begin{figure*}
\centering
\includegraphics[width=0.5\textwidth]{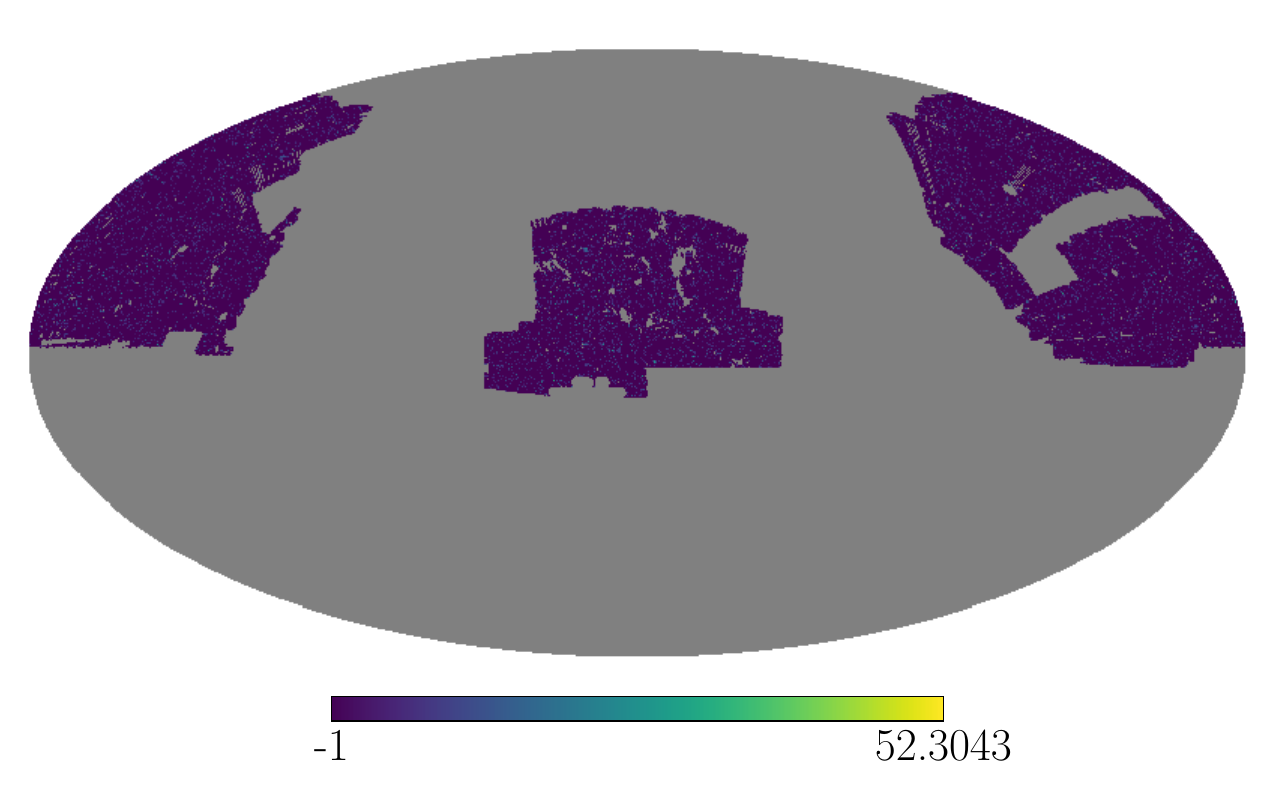}\includegraphics[width=0.5\textwidth]{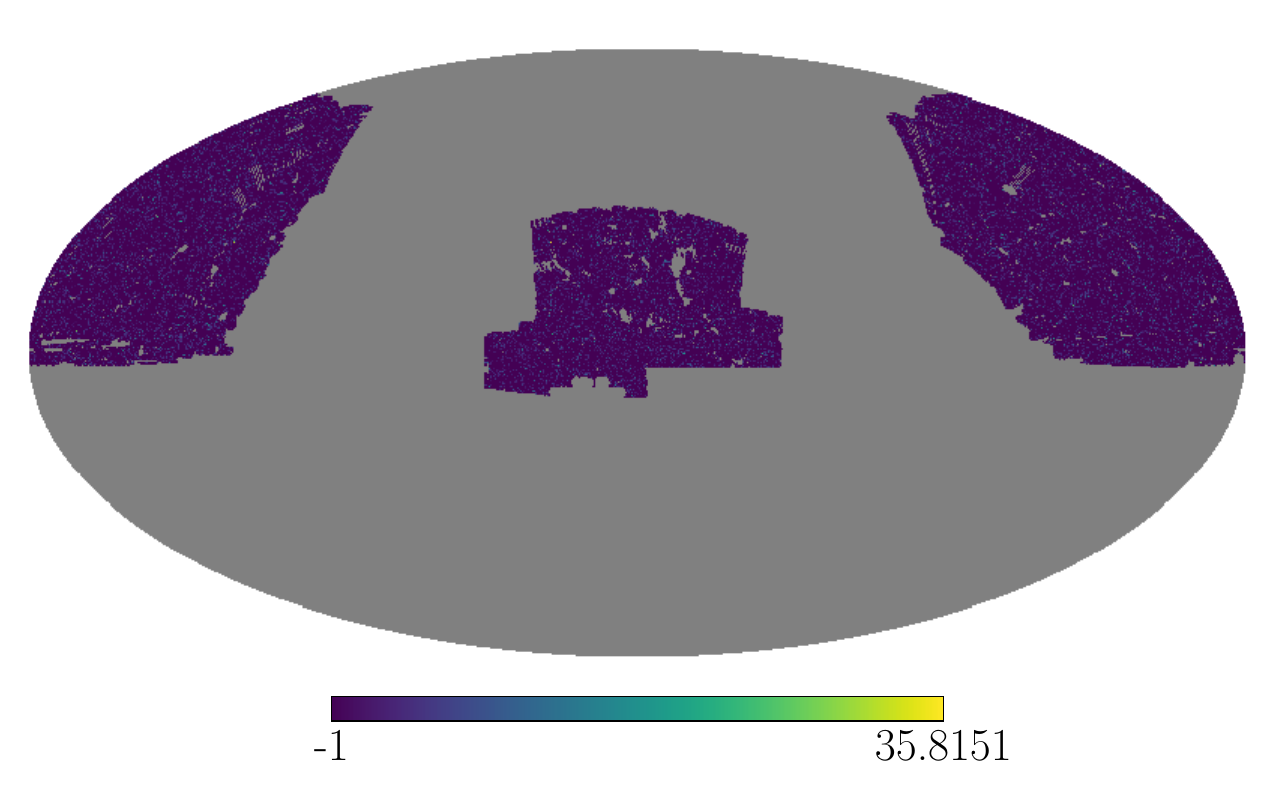}
\caption{LOWZ (left) and CMASS (right) galaxy overdensity maps in Celestial coordinates with  \(N_{\rm side}=1024\).}
\label{fig:overdensity_maps}
\end{figure*}

The resulting maps are shown in \autoref{fig:overdensity_maps}, where for simplicity we present only the cumulative galaxy angular distribution for each sample. As previously stated, the total sky areas correspond to \(f_{\rm sky}=0.2081\) for LOWZ and \(f_{\rm sky}=0.2416\) for CMASS, excluding the masked regions (shown in grey).

\section{Analysis}
\label{sec:analysis}
We shall now describe the procedure we follow to construct the likelihood function for our data, which is in turn used to derive measurements and constraints on the model parameters. Unless otherwise stated, we shall assume a Gaussian likelihood for the data. Since we are not interested in the overall normalisation of the posterior distribution, we can recast the analysis in terms of the minimisation of the chi-squared function,
\begin{equation}
     \chi^2\left(\{\theta_\alpha\}\right) = [\bm d - {\bm t}\left(\{\theta_\alpha\}\right)]^{\sf T}\,\mathsf C^{-1}\,[\bm d - {\bm t}\left(\{\theta_\alpha\}\right)]\;,
    \label{eq:chi2}
\end{equation}
where \(\{\theta_\alpha\}=\{\bs^i,\,\fs^i\}\) is the set of model parameters, which our theoretical prediction \(\bm t(\{\theta_\alpha\})\) depends upon, \(\bm d=\{d_a\}\) is the data vector, and \(\mathsf C=\{C_{ab}\}\) is the data covariance matrix. Here, \(i=1\ldots N_z\) runs over the number of redshift bins, \(N_z\), and \(a,b=1\ldots N_{\rm d}\) label the \(N_{\rm d}\) available data points.

We start from the definition of the data and theory vectors, and then move to the estimation of the covariance matrix. In our analysis, we adopt a pseudo power spectrum approach (also pseudo-\(C_\ell\)) \citep{Huterer_2001, Blake_2004,Ho_2012, Balaguera-Antolinez_2018}, consisting of projecting the observed field onto the celestial sphere, decomposing it into spherical harmonics, and then analysing statistically the coefficients of this decomposition after taking into account the incomplete sky coverage.

\subsection{Data and theory vectors}
The harmonic-space tomographic power spectrum can be estimated from the galaxy overdensity maps as\footnote{Throughout the text, estimators will be denoted by a wide hat, and pseudo-\(C_\ell\)'s with slashed letters.}
\begin{equation}
    \widehat{S^{\varDelta\varDelta}_{ij,\ell}}=\frac{1}{2\ell+1}\,\sum_{m=-\ell}^{\ell}\varDelta_{i,\ell m}\,\varDelta_{j,\ell m}-\frac{\delta^{\rm(K)}_{ij}}{\bar n_{\rm g}^i}\;,
    \label{eq:PCL}
\end{equation}
where \(\varDelta_{i,\ell m}\) are the harmonic coefficients of the pixelised overdensity map of \autoref{eq:overdensity}, and we have subtracted the shot-noise term (fourth column of \autoref{tab:specs}), which is diagonal in \(i-j\), \(\delta^{\rm(K)}\) being the Kronecker symbol.

\begin{figure*}
\centering
\includegraphics[width=0.5\textwidth]{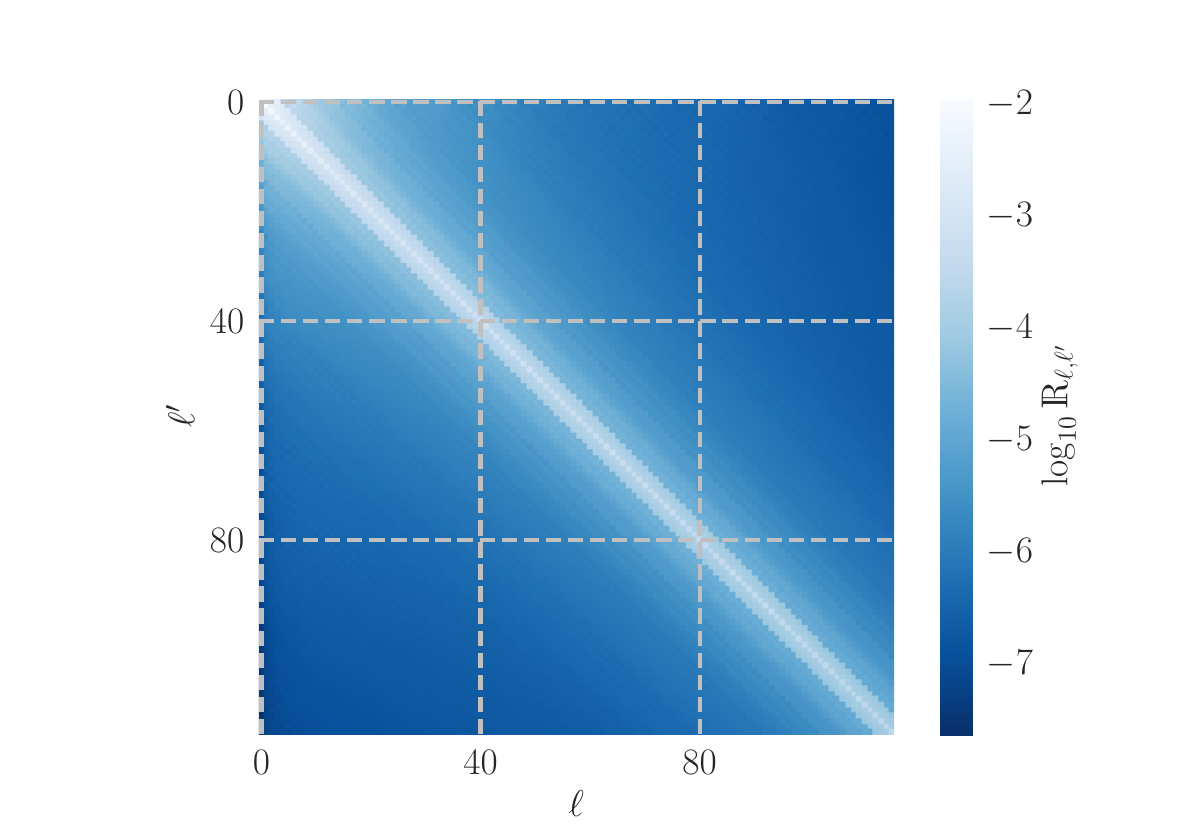}\includegraphics[width=0.5\textwidth]{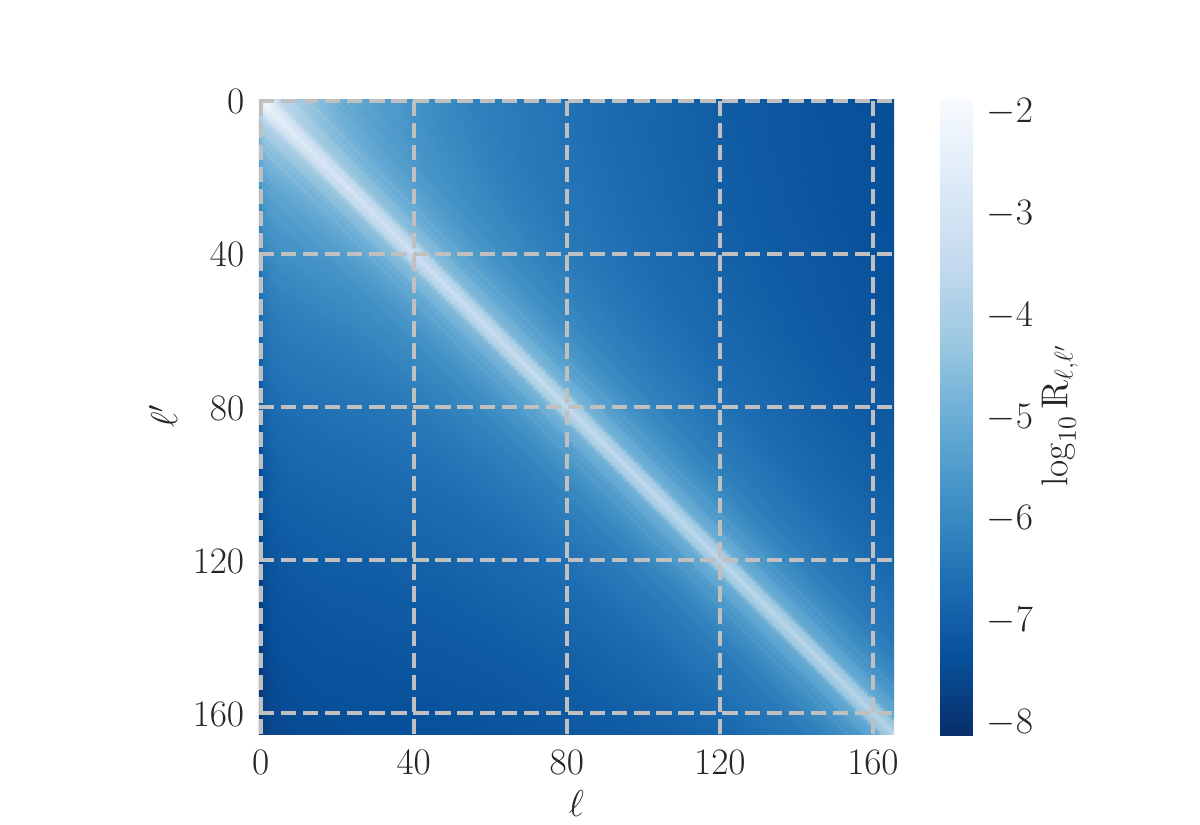}
\caption{The coupling matrices for LOWZ (left) and CMASS (right). The diagonal terms are dominant and the coupling between the different modes is given by the off-diagonal terms.}
\label{fig:coupling_matrix}
\end{figure*}

To relate the estimated power spectrum to the underlying one we need to account for the mask, which introduces a coupling between different multipoles. Since the unmasked field is statistically isotropic, it is related to the measured one via
\begin{equation}
\left\langle \widehat{S^{\varDelta\varDelta}_{ij,\ell}} \right\rangle = \sum_{\ell^\prime}\mathsf R_{\ell\ell^\prime}\,\widehat{S^{\varDelta\varDelta}_{ij,\ell^\prime}}\;,
    \label{eq:coupling1}
\end{equation}
where \(\mathsf R_{\ell\ell^\prime}\) is the coupling matrix. It is defined as
\begin{equation}
\mathsf R_{\ell\ell^\prime}=\frac{2\ell^\prime+1}{4\pi}\,\sum_{\ell^{\prime\prime}}(2\ell^{\prime\prime}+1)\,W_{\ell^{\prime\prime}}\begin{pmatrix}
\ell & \ell^\prime & \ell^{\prime\prime}\\
0 & 0 & 0
\end{pmatrix}^2\;,
    \label{eq:coupling2}
\end{equation}
with the matrix in square parentheses being the Wigner 3-j symbol and \(W_{\ell^{\prime\prime}}\) the harmonic-space power spectrum of the mask. The latter reads
\begin{equation}
W_{\ell}=\frac{1}{2\ell+1}\,\sum_{m=-\ell}^{\ell}\left|v_{\ell m}^{\phantom{\ast}}\right|^2\;,
    \label{eq:mask_spectra}
\end{equation}
where
\(v_{\ell m}\) are the coefficients of the harmonic decomposition of the binary mask \(v(\hat{\bm r})\). The coupling matrices for the LOWZ and CMASS masks are shown in the two panels of \autoref{fig:coupling_matrix}.

Generally, the direct inversion of the coupling matrix is not possible, because the loss of information due to the mask makes the coupling matrix singular and, therefore, the inversion ill-conditioned. One way to overcome this problem is to introduce bandpowers, which we do by utilising the public code \texttt{pymaster} \citep{Alonso2019}. Our bandpowers, \(s\), are a set of \(8\) multipoles with weights \(w^\ell_s\) normalised such that \(\sum_{\ell \in s}{w^\ell_s}=1\). Now, for the coupled pseudo-\(C_\ell\) in the \(s\)th bandpower we define
\begin{equation}
 \widehat{\slashed{S}^{\varDelta\varDelta}_{ij,s}}  = \sum_{\ell \in s}{w^\ell_s}\,\langle \widehat{S^{\varDelta\varDelta}_{ij,\ell}} \rangle\;,
    \label{eq:bandpower0}
\end{equation}
whose expectation value is then
\begin{equation}
\left\langle \widehat{\slashed{S}^{\varDelta\varDelta}_{ij,s}} \right\rangle = \sum_{\ell \in s}\,{w^\ell_s}\, \sum_{\ell^\prime}\mathsf R_{\ell\ell^\prime}\,\widehat{S^{\varDelta\varDelta}_{ij,\ell^\prime}}\;.
    \label{eq:bandpower1}
\end{equation}
By doing so, we implicitly assume that the true power spectrum is also a stepwise function, which is related to the bandpowers via
\begin{equation}
\widehat{S^{\varDelta\varDelta}_{ij,\ell}} = \sum_s \widehat{\slashed{S}^{\varDelta\varDelta}_{ij,s}}\,\Theta(\ell \in s)\;,
    \label{eq:bandpower2}
\end{equation}
with \(\Theta\) the Heaviside step function.

Finally, after inserting the above relation in \autoref{eq:bandpower0}, we can derive the unbiased estimator
\begin{equation}
\slashed{S}^{\varDelta\varDelta}_{ij,s} = \sum_{s^\prime} \mathsf M^{-1}_{ss^\prime}\, \widehat{\slashed{S}^{\varDelta\varDelta}_{ij,s^\prime}}\;,
    \label{eq:bandpower3}
\end{equation}
where \(\mathsf M\) is the binned coupling matrix, i.e.\
\begin{equation}
\mathsf M_{ss^\prime}=\sum_{\ell \in s}\sum_{\ell^\prime \in s^\prime}{w^\ell_s}\, \mathsf R_{\ell\ell^\prime}\;.
    \label{eq:bandpower4}
\end{equation}

Finally,  the theory spectra \(S^{\varDelta\varDelta}_{ij,\ell}\) should also be binned according to the bandpower choice in order to be compared with the data (for in general the theory curve is not a stepwise function). This translates into constructing
\begin{equation}
\slashed{S}^{\rm gg,th}_{ij,s}=\sum_{s^\prime} \mathsf M^{-1}_{ss^\prime}\,\sum_{\ell^\prime \in s^\prime}{w^{\ell^\prime}_{s^\prime}}\,\mathsf R_{\ell^\prime\ell}\,S^{\varDelta\varDelta}_{ij,\ell}\;,
    \label{eq:bandpower5}
\end{equation}
where we remind the reader that our theoretical model for \(S^{\varDelta\varDelta}_{ij,\ell}\) is defined in \autoref{eq:Cl_ours}.

\subsection{Covariance matrix}
Here, we shall describe the various approaches we have followed to estimate the data covariance matrix.

\subsubsection{Theory Covariance}
\label{subsec:Gauss}
The simplest scenario is to assume that the data covariance matrix is Gaussian, which then reads
\begin{equation}
\mathsf C^{ij,kl}_{ss^\prime}=
\frac{\delta^{\rm(K)}_{ss^\prime}}{(2s+1)\,\varDelta{s}\, f_{\rm sky}}\left(\slashed{C}^{\varDelta\varDelta}_{ik,s}\,\slashed{C}^{\varDelta\varDelta}_{jl,s}
+\slashed{C}^{\varDelta\varDelta}_{il,s}\,\slashed{C}^{\varDelta\varDelta}_{jk,s}\right)\;,\label{eq:gauss_covmat}
\end{equation}
with \(\varDelta{s=1}\) the multipole range in our bandwidth binning and
\begin{equation}
\slashed{C}^{\varDelta\varDelta}_{ij,s} = \slashed{S}^{\varDelta\varDelta}_{ij,s} + \frac{\delta^{\rm(K)}_{ij}}{\bar n^i_{\rm g}}\;.
    \label{eq:noise}
\end{equation}


\subsubsection{\texttt{PolSpice} Covariance}
\label{subsec:Pol}
The Gaussian covariance is, by construction, diagonal in multipoles, which is then reflected by the structure of \autoref{eq:gauss_covmat}---i.e.\ by the presence of the Kronecker symbol. However, as we have described above when introducing pseudo-$C_\ell$'s, convolution with the survey mask induces a coupling between modes. This, in turn, reflects onto the covariance matrix, and a way to estimate it is provided by the \texttt{PolSpice} package \citep{Chon2004}. This is described in detail in \citet{Efstathiou2004}, and the corresponding covariance matrix has following form
\begin{equation}
{\mathsf C}_{ss^\prime}=\mathsf M_{ss^\prime}^{-1}\,{\mathsf V}_{ss^\prime}\,\left[\left(\mathsf M^{-1}\right)_{ss^\prime}\right]^{\sf T}\;,
    \label{eq:pol_covmat}
\end{equation}
with
\begin{equation}
{\mathsf V}_{ss^\prime}=\frac{2\,\bm d_s\, \bm d_{s^\prime}\,\mathsf M_{ss^\prime}}{2\,s^\prime+1}\;.
    \label{eq:V_mat}
\end{equation}
We remind the reader that \(\mathsf M_{ss^\prime}\) is the binned coupling matrix, given in \autoref{eq:bandpower4}, and $\bm d_s$ the data vector corresponding to bandpower \(s\).


\subsubsection{Mocks Covariance}
\label{subsec:Mock}
The most realistic way to obtain a realistic data covariance matrix is through simulating the data itself and extracting its covariance from the sample. The BOSS collaboration provides us with simulated galaxy catalogues, namely the \texttt{PATCHY} \citep{Kitaura2016} and the \texttt{QPM} \citep{Wang2017,White2013} mocks. They are constructed assuming a fixed cosmological model and can be used to estimate accurate covariance matrices since they include various observational and systematic effects. In this work, we have selected the \texttt{QPM} mocks to calculate the covariance matrix for the LOWZ sample and the \texttt{PATCHY} mocks for the CMASS sample. This decision is made due to the different redshift ranges covered by the simulated catalogues. Indeed, the \texttt{PATCHY} mocks contain galaxies at redshifts \(0.2<z<0.75\) whilst our selected LOWZ sample starts from \(z>0.15\). The selection of two different sets of mocks is a further test to validate our analysis on top of the consistency checks that we make in appendix~\ref{appendix}.

The resulting covariance matrix is then
\begin{equation}
{\mathsf C}_{ss^\prime}=\frac{1}{N_{\rm m}-1}\,\sum_{m=1}^{N_{\rm m}}\left(\bm d_s^m-\bar{\bm d}_s\right)\,\left(\bm d_{s^\prime}^m-\bar{\bm d}_{s^\prime}\right)^{\sf T}\;,
    \label{eq:mock_covmat}
\end{equation}
with \(m=1\ldots N_{\rm m}\) running through the \(N_{\rm m}=1000(2048)\) available \texttt{QPM}(\texttt{PATCHY}) mocks, \(\bm d_s^m=\{\bm{\slashed{S}}^{\varDelta\varDelta,m}_{ij,s}\}\) being the mock data vector, and 
\begin{equation}
\bar{\bm d}_s^m=\frac{1}{N_{\rm m}}\,\sum_{m=1}^{N_{\rm m}}\bm d_s^m\;.
    \label{eq:mock_average}
\end{equation}

Some extra care has to be taken when the covariance matrix is estimated using simulations, and the final likelihood should be corrected accordingly. This is because the inverse of the covariance matrix derived from simulations can be a biased estimator of the inverse of the true covariance matrix \citep{Hartlap}. Here, we present two methods that account for this correction:
\begin{enumerate}
    \item The former was first proposed by \citet{Hartlap} and it based on the following reasoning. Whilst the covariance matrix inferred from simulations can be an unbiased estimator of the true covariance \(\mathsf C\), its inverse (entering the \(\chi^2\) in \autoref{eq:chi2}) is not, and therefore should be rescaled according to \citep{Anderson2004}
\begin{equation}
{\mathsf C}^{-1} \rightarrow \frac{N_{\rm m}-N_{\rm d}-2}{N_{\rm m}-2}\,{\mathsf C}^{-1}\;.
    \label{eq:Anderson_Hartlap}
\end{equation}
Note that by doing so the assumption of a Gaussian likelihood can be maintained.
    \item The latter has been proposed by \citet{SH2016}, where the authors state that the Gaussian likelihood should be replaced by the Student's \(t\)-distribution,
\begin{equation}
\mathcal{L}\propto \left[1+\frac{ \chi^2}{N_{\rm m}-1}\right]^{-N_{\rm m}/2}\;,
    \label{eq:SH_like}
\end{equation}
replacing as well the Gaussian covariance in the \(\chi^2\) of \autoref{eq:chi2} with \autoref{eq:mock_covmat}. The proportionality in \autoref{eq:SH_const} is set by
\begin{equation}
\frac{{\Gamma} \left(N_{\rm m}/2\right )\,[\text{det}({\mathsf C})]^{-1/2}}{[\pi\,(N_{\rm m}-1)]^{N_{\rm d}/2}\,\Gamma \left[(N_{\rm m}-N_{\rm d})/2\right]}\;,
    \label{eq:SH_const}
\end{equation}
with \(\Gamma\) the Euler Gamma function and assuming \(N_{\rm m}>N_{\rm d}\).
\end{enumerate}

We note that the results after implementing both methods are equivalent, and therefore we present results of the latter correction alone.

\section{Results}
\label{sec:result}
For the parameter estimation in this work we use the Bayesian-based sampler \texttt{emcee} \citep{ForemanMackey2013}. Before we proceed with the presentation of our results, we should note that throughout our analysis, we consider in the data vector and the covariance matrix only the equal bin correlations. In other words, in the data vector we only include auto-bin spectra, $\slashed{S}^{\varDelta\varDelta}_{ii,s}$; and in the covariance matrix, we keep only auto-bin-pair ($\{i-i$, $i-i\}$ and $\{j-j$, $j-j\}$) and cross-bin-pair ($\{i-i$, $j-j\}$ and $\{j-j$, $i-i\}$) terms---thus ignoring combinations like $\{i-j$, $i-j\}$ or $\{i-j$, $j-j\}$.
We do so because these contributions are not expected to encode much information on the growth rate of structures. This is due to the fact that the redshift bins do not overlap, as shown in the top panel of \autoref{fig:Nz_and_comparison_2}. Although it is true that RSD effectively induce cross-bin correlations \citep[see][for a heuristic explanation using not the exact solution as in this paper but rather the Limber approximation]{2019MNRAS.489.3385T}, we can safely neglect its effect in the analysis, as it will be shown later (see \autoref{appendix}). Note that this choice is similar to Fourier-space analyses, where cross-correlations among redshift bins are not considered.


\begin{figure}
\centering
\includegraphics[width=\columnwidth]{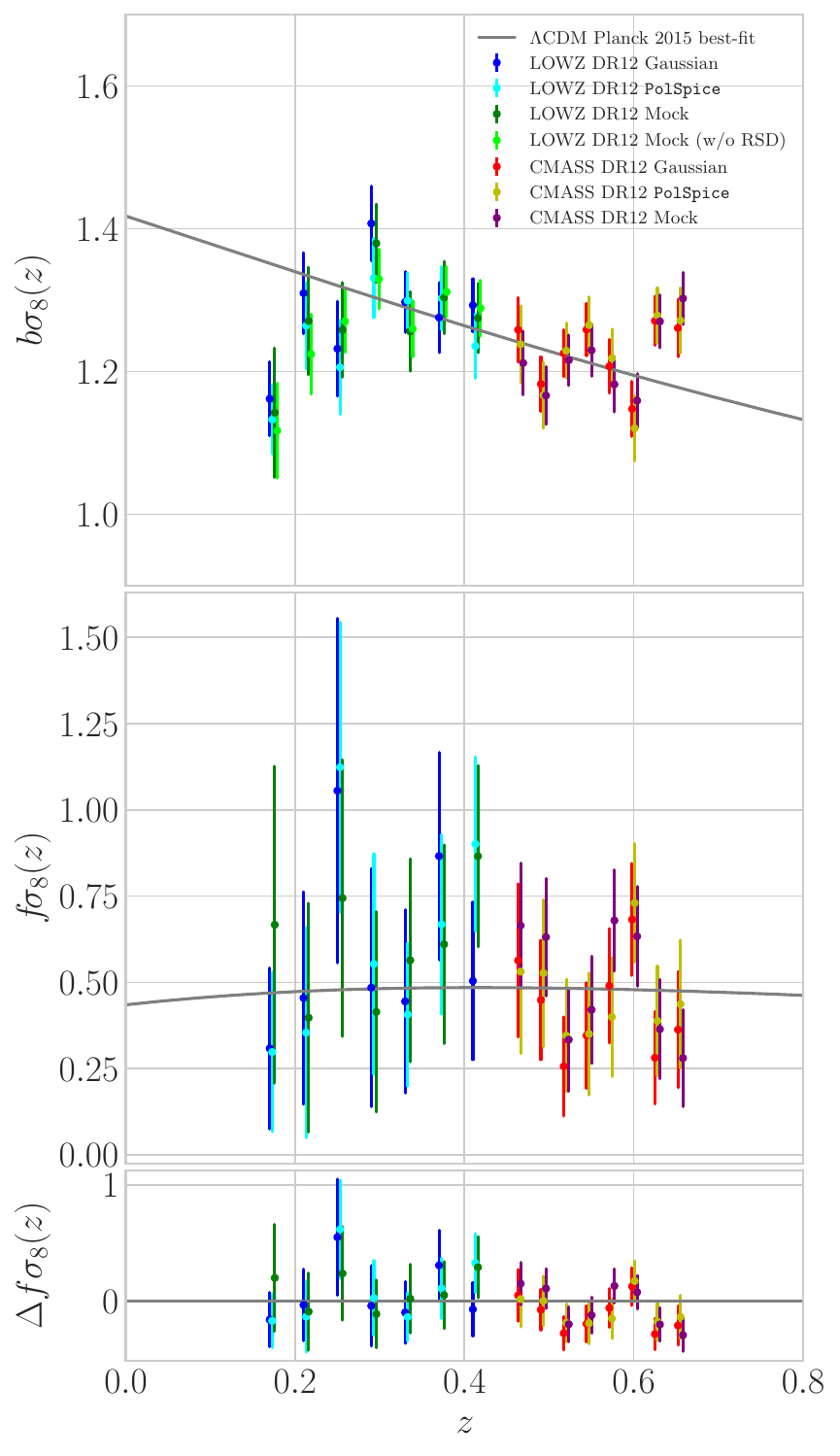}
\caption{Estimated means and \(68\%\) C.L.\ intervals on \(\bs^i\) (top panel) and \(\fs^i\) (central panel) as a function of redshift, for the bin choices of LOWZ and CMASS. Results for LOWZ with the Gaussian covariance, the \texttt{PolSpice} covariance, and the mock covariance are respectively shown with blue, cyan, and green, whilst for CMASS they instead are rendered in red, gold, and purple. In light green (top panel) we show the \(\bs^i\) results for LOWZ and the mocks covariance neglecting the RSD (see text for details). Note that constraints on each redshift bin from the different covariance estimations are plotted with a slight shift of \(0.003\) from one another, to enhance readability. The solid grey curves are the \(\bs(z)\) (top panel) and \(\fs(z)\) (central panel) prediction for our reference cosmology \citep{Ade2015}. The bottom panel shows the residuals on \(\fs(z)\).}
\label{fig:results}
\end{figure}

In \autoref{fig:results}, we show the constraints on the parameter set \({\{\theta_\alpha\}}=\{\bs^i,\fs^i\}\) for the three covariance estimates described in \autoref{subsec:Gauss}, \autoref{subsec:Pol}, and \autoref{subsec:Mock}. In particular, means along with their 68\% confidence level (C.L.) intervals are presented for each case in \autoref{tab:results_lowz_cmass}. This is the main result of this paper, and we shall now discuss it in detail.
\begin{table}
\centering
\caption{Means and corresponding \(68\%\) C.L.\ intervals on \(\bs^i\) and \(\fs^i\) in the \(i\)th redshift bin for the LOWZ (upper table with \(i=1\ldots 7\)) and CMASS (lower table with \(i=1\ldots 8\)) sample using the Gaussian, the \texttt{PolSpice}, and the mock covariance matrix (see \autoref{fig:results}).}
\setlength{\tabcolsep}{2.3pt}
\renewcommand{\arraystretch}{0.9}
\begin{tabularx}{\columnwidth}{X c c c}
    \hline
   & Gaussian cov.\ mat.\ & \texttt{PolSpice} cov.\ mat.\ & Mock cov.\ mat.\ \\
    \hline
    \hline
    $\bs^1$ &$1.162\pm0.052$ & $1.133\pm0.049$ & $1.142\pm0.091$ \\
    $\bs^2$ & $1.309\pm0.057$ & $1.264\pm0.060$ & $1.271\pm 0.075$ \\
        $\bs^3$ & $1.232\pm0.066$ & $1.206\pm0.066$ & $1.258\pm0.066$ \\
            $\bs^4$ & $1.407\pm0.052$ & $1.331\pm0.055$ & $1.379\pm0.055$ \\
            $\bs^5$ & $1.297\pm0.043$ & $1.299\pm0.039$ & $1.256\pm0.055$ \\
            $\bs^6$ & $1.275\pm0.049$ & $1.302\pm0.044$ & $1.303\pm0.051$ \\
            $\bs^7$ & $1.292\pm0.037$ & $1.235\pm0.045$ & $1.275\pm0.048$ \\            
    \hline

    $\fs^1$ & $0.308\pm 0.233$ & $0.298\pm0.231$ & $0.667\pm0.459$ \\
    $\fs^2$ & $0.454\pm0.307$ & $0.354\pm0.305$ & $0.398\pm 0.332$ \\
        $\fs^3$ & $1.055\pm0.499$ & $1.123\pm 0.420$ & $0.744\pm0.401$ \\
            $\fs^4$ & $0.484\pm0.345$ & $0.553\pm0.319$ & $0.414\pm 0.290$ \\
            $\fs^5$ & $0.445\pm 0.266$ & $0.406\pm 0.207$ & $0.564\pm 0.294$ \\
            $\fs^6$ & $0.866\pm 0.301$ & $0.668\pm 0.260$ & $0.610\pm 0.288$ \\
            $\fs^7$ & $0.504\pm 0.228$ & $0.901\pm 0.253$ & $0.866\pm 0.263$ \\            
    \hline
    \hline

    $\bs^1$ &$1.258\pm0.045$ & $1.238\pm0.054$ & $1.212\pm0.047$ \\
    $\bs^2$ & $1.182\pm0.038$ & $1.167\pm0.046$ & $1.166\pm 0.040$ \\
        $\bs^3$ & $1.226\pm0.033$ & $1.229\pm0.039$ & $1.215\pm0.035$ \\
            $\bs^4$ & $1.258\pm0.036$ & $1.265\pm0.039$ & $1.230\pm0.036$ \\
            $\bs^5$ & $1.207\pm0.038$ & $1.219\pm0.041$ & $1.182\pm0.038$ \\
            $\bs^6$ & $1.148\pm0.039$ & $1.121\pm0.045$ & $1.159\pm0.037$ \\
            $\bs^7$ & $1.271\pm0.034$ & $1.279\pm0.038$ & $1.270\pm0.037$ \\
            $\bs^8$ & $1.261\pm0.040$ & $1.271\pm0.046$ & $1.302\pm0.036$ \\
    \hline

    $\fs^1$ & $0.564\pm 0.222$ & $0.531\pm0.237$ & $0.665\pm0.181$ \\
    $\fs^2$ & $0.448\pm0.173$ & $0.527\pm0.213$ & $0.631\pm 0.170$ \\
        $\fs^3$ & $0.256\pm0.144$ & $0.346\pm 0.163$ & $0.334\pm0.150$ \\
            $\fs^4$ & $0.346\pm0.152$ & $0.351\pm0.177$ & $0.420\pm 0.156$ \\
            $\fs^5$ & $0.490\pm 0.166$ & $0.398\pm 0.172$ & $0.680\pm 0.147$ \\
            $\fs^6$ & $0.682\pm 0.162$ & $0.730\pm 0.172$ & $0.633\pm 0.144$ \\
            $\fs^7$ & $0.282\pm 0.134$ & $0.389\pm 0.159$ & $0.364\pm 0.144$ \\   
            $\fs^8$ & $0.363\pm 0.168$ & $0.437\pm 0.185$ & $0.281\pm 0.141$ \\ 
    \hline
\end{tabularx}
\label{tab:results_lowz_cmass}
\end{table}

First, the results assuming the Gaussian covariance are shown in blue for LOWZ and red for CMASS for the \bs\ parameters in the top panel of \autoref{fig:results} and the \fs\ parameters in the central panel of \autoref{fig:results}. Note that we validate these results against potential systematic effects after performing a series of consistency tests, as described in \autoref{appendix}.

Regarding the constraints on clustering bias in the top panel, we see the \(\bs(z)\) measurements to be around \(1.3\) for LOWZ and \(1.2\) for CMASS, in agreement with the literature on measurements from BOSS DR12. For example, our findings are consistent with those obtained by \citet[][see Figure\ 7]{Salazar2016}, where they have assumed tomography not in harmonic space but using the two-point correlation function. The solid, grey curve represents the galaxy bias functional form described in \citet{Salazar2016}, further multiplied by $\sigma_8(z)$, against which our measurements show a good qualitative agreement.

Now, we turn our attention to the results of \(\fs(z)\) in the central panel. Here, we also show the theoretical prediction for the concordance \lcdm\ model from Planck as a solid, grey curve. Our measurements are randomly scattered around it and are close to this prediction, indicating that they are in excellent agreement. This can be easily appreciated by looking at the residuals, shown in the bottom panel of \autoref{fig:results}. 


At this point, it is worth commenting the fact that our method provides measured errors on \fs\ larger than those estimated from traditional methods in Fourier space and configuration space. The main point to raise is that we are here presenting a proof-of-concept analysis, for which we prefer to be conservative and stick to strictly linear scales. Instead, e.g.\ in the aforementioned \citet{Salazar2016}, the authors follow previous BOSS analysis set-ups like \citet{2017MNRAS.464.1640S} or \citet{2017MNRAS.467.2085G} and implement a perturbation theory approach including up to one-loop corrections.

Moreover, our thinner-binning tomographic approach allows us to track more finely the growth rate as a function of redshift. Finally, note that forthcoming experiments such as the European Space Agency's Euclid mission \citep{Laureijs2011,Amendola2013,Amendola2016}, the Dark Energy Spectroscopic Instrument \citep{DESIcollab2016}, and SKA Observatory's radio-telescopes \citep{Maartens2015,Abdalla2015,SKA1_2018} will push to much deeper redshift, where the reach of linear theory is larger, thus allowing for a significant improvement in the precision on the measurements with the method we have presented here.

Now, let us focus on the clustering and growth rate results using the \texttt{PolSpice} covariance. These measurements are shown in \autoref{fig:results} with cyan for LOWZ and gold for CMASS. We can appreciate that these measurements are very consistent and similar with those obtained with the Gaussian covariance, meaning that the mode coupling induced by the mask does not significantly increase the noise.

Finally, we calculate the constraints on \bs\ and \fs\ with the mock covariance as estimated from \texttt{QPM} mocks for LOWZ shown with green, and from \texttt{PATCHY} mocks for CMASS shown with purple in \autoref{fig:results}. Note that, in the absence of a reliable theoretical model for the data covariance, using mock data to estimate the covariance represents the most agnostic approach to the analysis of the data \citep[see also][]{2022MNRAS.510.3207P}. Again, we can appreciate that the agreement of data from the mock covariance with those form the \texttt{PolSpice} and the Gaussian ones is quite good. Interestingly, the agreement worsens at lower redshift, where we do in fact expect non-Gaussian terms in the covariance matrix to be more important.

At this point it is instructive to note that some slight changes between the results from the three covariance matrices (although the overall agreement is quite good and we therefore decide to present all of them) are expected due to the by construction differences from their definitions (see \autoref{subsec:Gauss}, \autoref{subsec:Pol} and \autoref{subsec:Mock}). The Gaussian covariance is an approximation of the true covariance that is not precise in the presence of non-linearities and also does not account for the mode coupling due to mixing matrix. Nevertheless, it is worth noting that it shows a good agreement with the other, more sophisticated method. This confirms the finding of \citet{Loureiro2019}.

The \texttt{PolSpice} covariance, on the other hand, incorporates the mode coupling but does not take into account the cross-bin-pair combinations in the covariance ($\{i-i$, $j-j\}$ and $\{j-j$, $i-i\}$). Summarising, we stress that the mocks covariance is expected to give the precise estimate for the true covariance matrix as it reaches an infinite number of simulations.


\begin{table}
\centering
\caption{The reduced \(\chi^2\) (\(\chi^2/{\rm d.o.f.}\)) for LOWZ and CMASS using the three covariance matrices}
\setlength{\tabcolsep}{2.3pt}
\renewcommand{\arraystretch}{0.9}
\begin{tabularx}{\columnwidth}{X c c c}
    \hline
   & Gaussian cov.\ mat.\ & \texttt{PolSpice} cov.\ mat.\ & Mock cov.\ mat.\ \\
    \hline
    \hline
    LOWZ & $2.980$ & $2.225$ & $2.218$ \\
    CMASS & $2.060$ & $1.244$ & $1.620$ \\
    \hline
\end{tabularx}
\label{tab:reduced_chi2}
\end{table}

To conclude, we quote in \autoref{tab:reduced_chi2} the values of the reduced \(\chi^2\) for the two samples and the three covariance matrices. These numbers, all of order of a few, give us a qualitative confirmation that the fitting template and the estimated covariance matrices correctly capture the data. In addition to that, the more sophisticated methods for the covariance matrix (\texttt{PolSpice} and mocks) perform better, as expected, compared to the Gaussian estimate for both LOWZ and CMASS.

Note that, however, there is a trend for a higher reduced $\chi^2$ value for LOWZ compared to CMASS. This, in fact, could be due to the less constraining power of LOWZ (see \autoref{fig:results}) which is impacting the goodness-of-fit in the following sense. Low-redshift measurements, like LOWZ, are much more sensitive to non-linear evolution and since our method works only on linear scales, it might be that the constraining power in the data is not enough to put competitive constraints on the growth rate and it would rather prefer a simpler model, w/o RSD.

Thus, we set up a run for LOWZ and the mocks covariance matrix in which we neglect altogether the growth rate contribution, removing the second and the third terms of \autoref{eq:Cl_ours} and put constraints only on the \bs\ parameters. To account for the loss of information on the \fs\ parameters, we also add a global nuisance parameter in the form of an Alcock-Paczy\'nski parameter, $\alpha_\perp$. The formulation of this parameter in a template fitting relation like \autoref{eq:Cl_ours} is simple and is presented in \cite{Stef22}. Now the reduced \(\chi^2\) is lowered to the satisfactory value of \(1.586\). These \bs\ measurements, shown with light green colour in \autoref{fig:results}, are consistent within $1\sigma$ with the other scenarios. Lastly, we point the reader to what follows for more detailed consistency tests in which we also present the \fs\ constraints with LOWZ for the sake of complicity and comparison with CMASS.

\section{Consistency tests}
\label{appendix}
In this section, we run a battery of tests to validate our pipeline and to check for possible systematic effects. For these tests we use the Gaussian covariance described in \autoref{subsec:Gauss} unless otherwise stated. All the results are presented in \autoref{fig:checks_fig2}, \autoref{fig:checks_fig3_fig1} and \autoref{fig:residuals}.

\subsection{Mock data vector}
\begin{figure}
\centering
\includegraphics[width=\columnwidth]{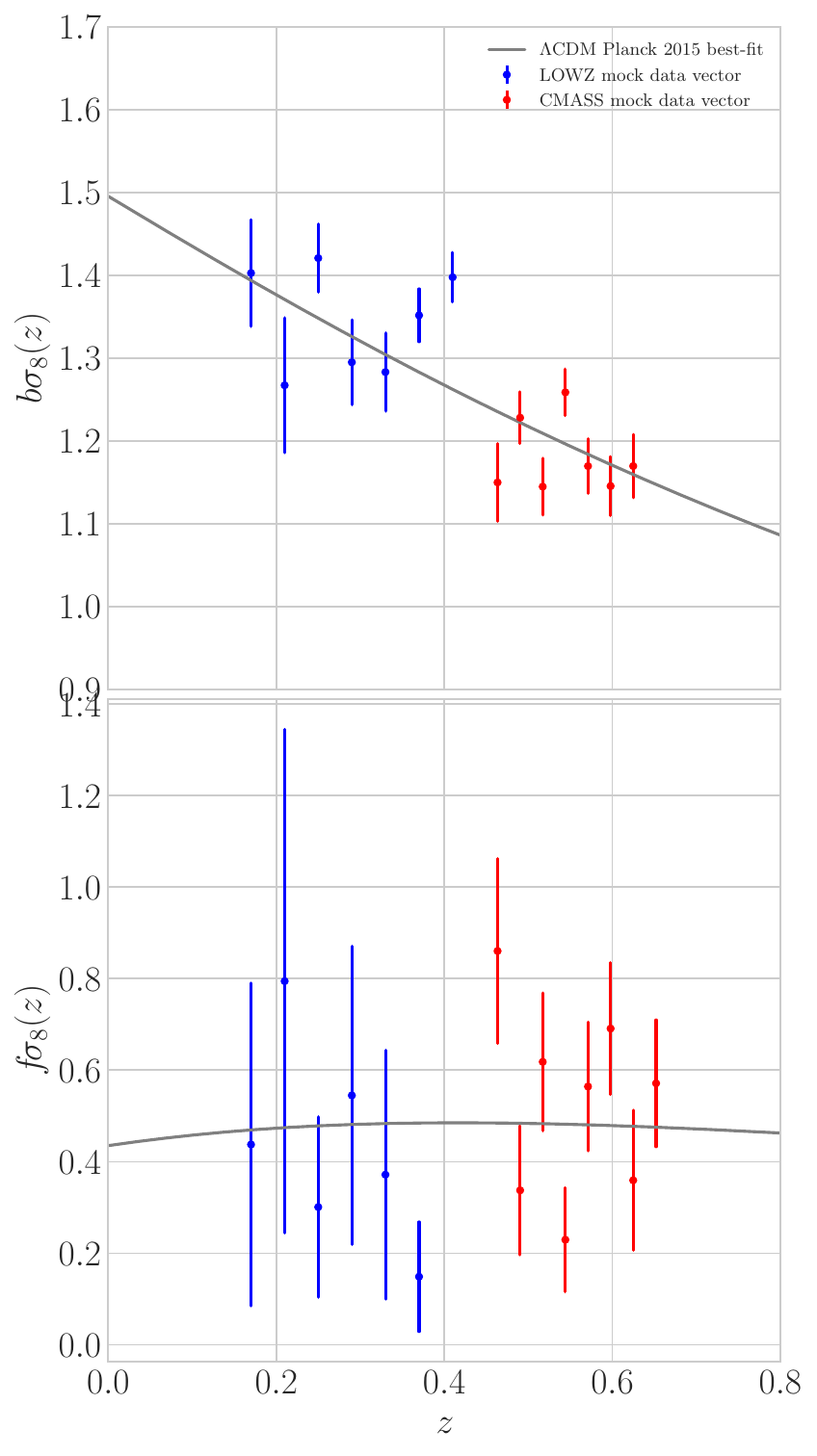}
\caption{Similar to \autoref{fig:results}, but for the mock data vector. Results on \bs\ (left panel) and \fs\ (right panel) are shown for LOWZ and CMASS in blue and red, respectively.}
\label{fig:checks_fig2}
\end{figure}

\begin{figure*}
\centering
\includegraphics[width=\columnwidth]{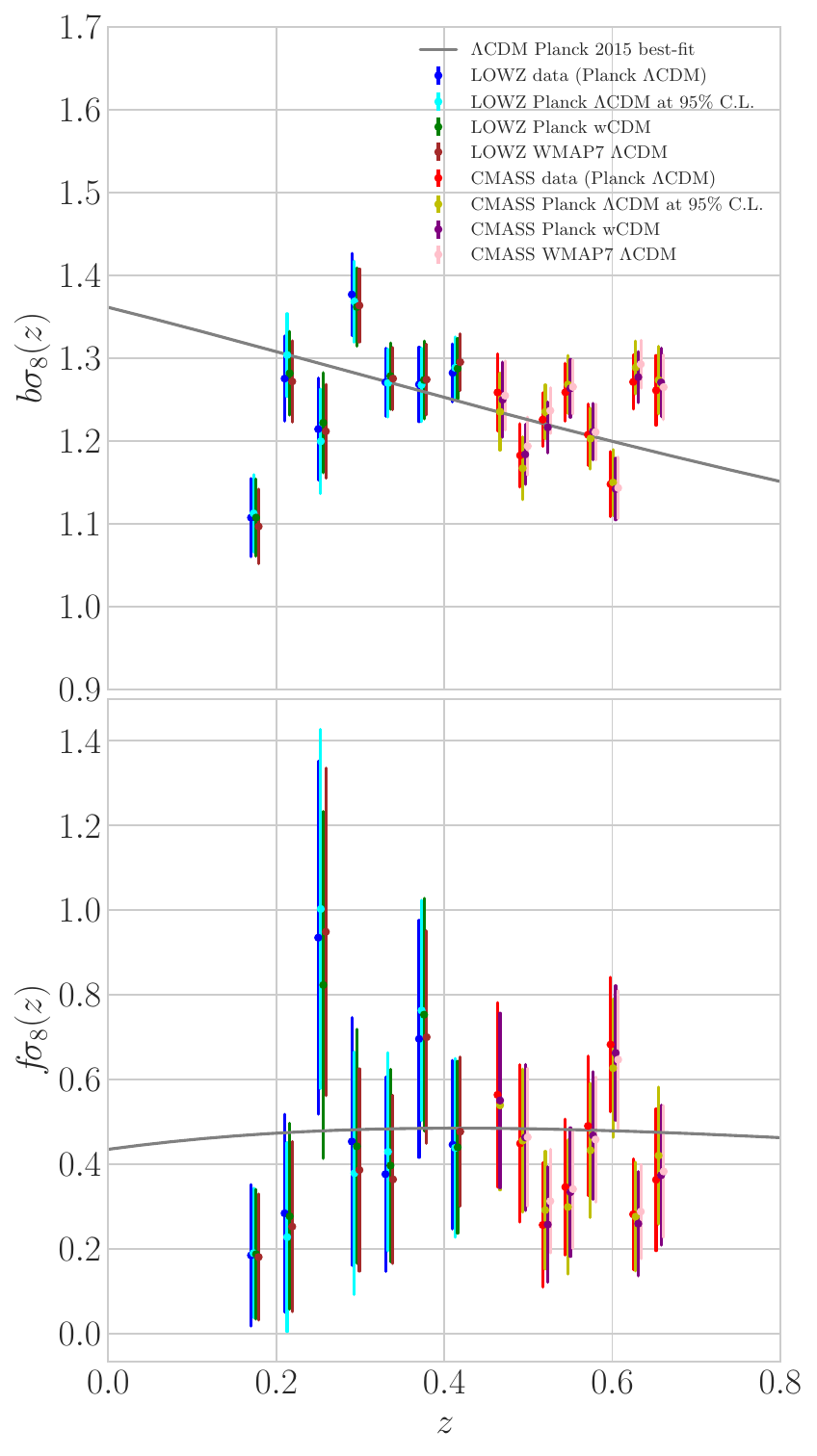}
\includegraphics[width=0.472\textwidth]{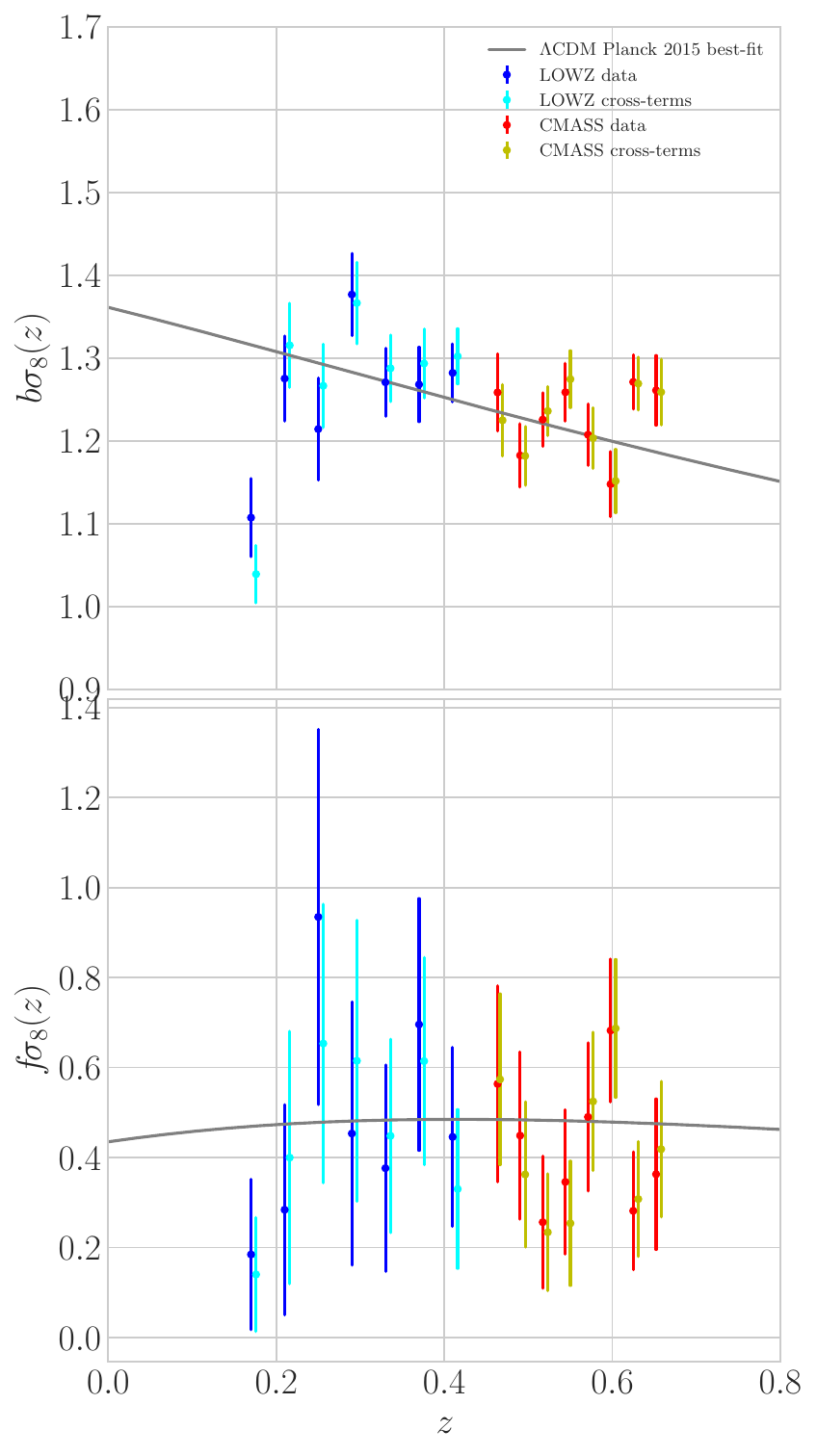}
\caption{\textit{Left:} Same as \autoref{fig:results}, but for the different fiducial cosmologies considered in \autoref{appendix}. We show the \(68\%\) C.L.\ intervals (error bars) and the means (circles) for \bs\ (top panel) and \fs\ (bottom panel) as a function of redshift for the bin choices of LOWZ and CMASS. Results for LOWZ with the original analysis assuming the Planck best-fit \lcdm\ cosmology, the Planck \lcdm\ cosmology at the 95\% C.L of the best-fit, the Planck best-fit \(w\)CDM cosmology and the WMAP7 \lcdm\ cosmology are respectively rendered in blue, cyan, green and brown, whilst for CMASS they are in red, gold, purple and pink. Again, the constraints on each redshift bin from the different tests are overplotted with \(0.003\) for clarity, and the solid grey curves are the \lcdm\ \citep{Ade2015}.\textit{Right}: In the same spirit as with the left panel but now for the cross-terms test that we have made (see \autoref{appendix}). Results for LOWZ with the original analysis and all the bin correlations (including all the cross-terms) respectively rendered in blue and cyan, whilst for CMASS they are in red and gold.}
\label{fig:checks_fig3_fig1}
\end{figure*}

As a first test, we want to replace the data vector \(\bm d_s\) with a mock data vector \(\bm d_s^m\) randomly chosen from the \texttt{QPM} or \texttt{PATCHY} sets of available mocks for LOWZ or CMASS. If our pipeline does not suffer from systematic effects, the analysis using the mock vector is expected to produce results with comparable constraining power for the same samples with those of the real data vector, but the points will be scattered differently around the theory prediction. As shown in \autoref{fig:checks_fig2}, this is indeed the case, with the mock data vector performing similarly to the real data. 
In particular, values with the real data results for the \bs\ (blue for mock in \autoref{fig:checks_fig2} and blue for real data in the right panel of \autoref{fig:checks_fig3_fig1} ) and for \fs\ (red for mock in \autoref{fig:checks_fig2} and red for real data in the right panel of \autoref{fig:checks_fig3_fig1}) all randomly scattered around the \lcdm\ prediction (solid grey line).

\subsection{Dependence on fiducial cosmology}
As discussed in \autoref{sec:methodology}, our method allows for almost model-independent measurements of the bias and growth of galaxies directly in harmonic space. However, we still have to assume a cosmology to compute the various ingredients of \autoref{eq:factorised_delta-delta}-\ref{eq:factorised_delta-v}.

In order to validate our working hypothesis, we here change the underlying fiducial cosmology that is assumed for the theory spectra \(T^{\delta\delta}_{ij,\ell}\), \(T^{VV}_{ij,\ell}\), and \(T^{\delta V}_{ij,\ell}\) calculated in \texttt{CLASS}. In particular, we first vary the cosmological parameters at the edge of their \(95\%\) C.L.\ for Planck \lcdm\ best-fit values of \citep{Ade2015}, namely \(\Omega_{\rm b}{=0.0621}\), \(\Omega_{\rm c}{=0.2718}\), and \(H_{0}{=68.67\,\mathrm{km\,s^{-1}\,Mpc^{-1}}}\). Furthermore, we perform the analysis assuming the best-fit values of the Planck \(w\)CDM cosmology of \citep{Ade2015} with \(\Omega_{\rm b}=0.04802\), \(\Omega_{\rm c}=0.2568\), \(H_0=68.1\,\mathrm{km\,s^{-1}\,Mpc^{-1}}\) and \(w=-1.019\). Finally we consider a more significantly different cosmology as the \lcdm\ best-fit from WMAP7 \citep{Komatsu2011}, with \(\Omega_{\rm b}=0.045\), \(\Omega{\rm c}=0.227\) and \(H_0=70.4\,\mathrm{km\,s^{-1}\,Mpc^{-1}}\).

Constraints for these three cosmologies are respectively shown in the left panel of \autoref{fig:checks_fig3_fig1}: cyan, green, and brown for LOWZ; and gold, purple, and pink for CMASS. The reconstructed values of \(\bs^i\) and \(\fs^i\) are well within the \(68\%\) C.L.\ intervals amongst all cosmologies (shown with blue and red), implying that the most part of the cosmological information is contained in \(\bs(z)\) and \(\fs(z)\) themselves.

\subsection{Including all correlations in the covariance matrix and data vector}
A further test for our pipeline is to check the effect of neglecting cross-bin terms in the data vector---i.e.\ $\slashed{S}^{\varDelta\varDelta}_{ij,s}$ with $i\ne j$---and the corresponding entries in the covariance matrix---namely, terms like $\{i-j$, $i-j\}$, $\{i-j$, $j-j\}$, $\{j-i$, $i-j\}$ etc. Basically, we use the full covariance matrix as defined in \autoref{eq:gauss_covmat}. Marginalised constraints presented in right panel of \autoref{fig:checks_fig3_fig1} illustrated in brown for LOWZ and pink for CMASS, respectively, show a satisfactory consistency with the default case where we do not consider the cross-bin-no-pair correlations at all.

\subsection{Residual distribution}
Finally, we perform the following test to validate our derived cosmological measurements \citep[see also][]{Loureiro2019}. This test is described as follows. In the case that we have a diagonal covariance matrix, we are able to construct a vector
\begin{equation}
\bm R = \mathsf O^{-1}\, [\bm d-\bm t]\;,
    \label{eq:residuals1}
\end{equation}
which is the normalised residuals, with \(\mathsf O\) the diagonal matrix constituted of the square root of the variances, \(\bm d\) the data vector, and \(\bm t\) the theory vector. Following e.g.\ \citet{andrae2010dos}, we note that the residuals are described by a standard normal distribution if the model given by \(\bm t\) represents the data and the errors are known exactly. Nonetheless, the errors are usually estimated from the limited number of samples provided by the experiments and the residual distribution is therefore given by the Student's \(t\)-distribution, which approaches the standardised Gaussian as the number of samples increases. Hence, if the residuals are described by a standard distribution, we can either infer that either we found the correct model, or that the data are not good enough to show any model preference. On the other hand, the residual distribution deviates from Gaussianity, then the model is ruled out.

In the realistic scenario, we know that the errors are correlated and the covariance matrix is no more diagonal. However, we can diagonalise it as
\begin{equation}
\mathsf C = \mathsf Q\, \mathsf Y\, \mathsf Q^{-1}\;,
    \label{eq:residuals2}
\end{equation}
with \(\mathsf Q\) the eigenvector matrix and \(\mathsf Y\) a diagonal matrix constructed from the eigenvalues of the original covariance matrix, \(\mathsf C\). By treating the new diagonal covariance matrix as \(\mathsf Y\), the residual distribution now reads
\begin{equation}
\bm R = \mathsf O^{-1}\, \mathsf Q^{-1} \,[\bm d - \bm t]\;,
    \label{eq:residuals}
\end{equation}
where \(\mathsf O\) contains the square roots of the eigenvalues.
\begin{figure*}
\centering
\includegraphics[width=0.5\textwidth]{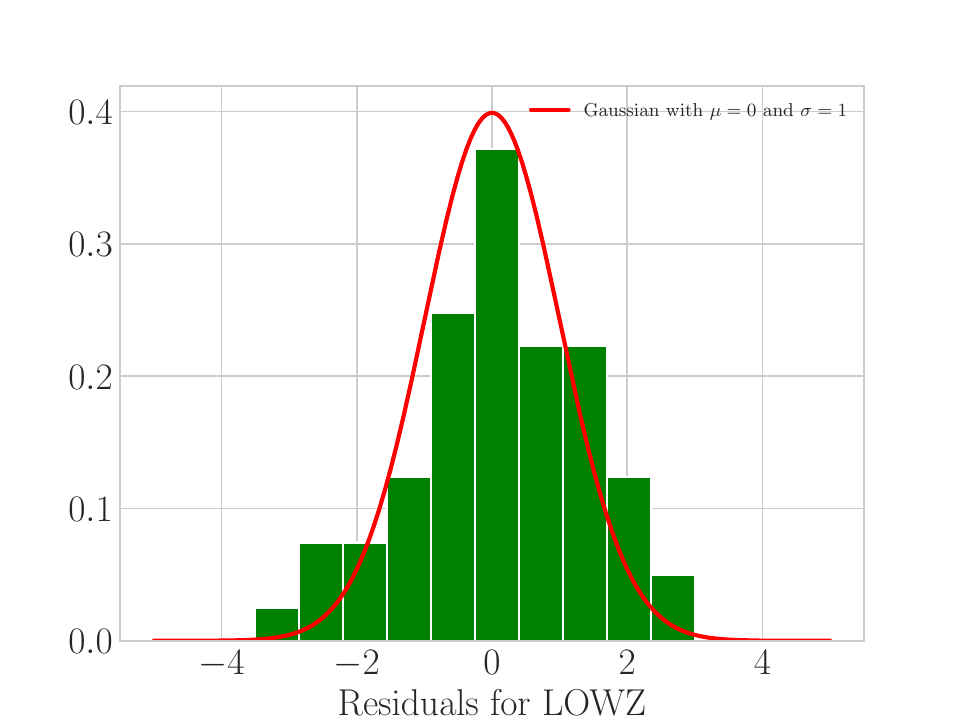}\includegraphics[width=0.5\textwidth]{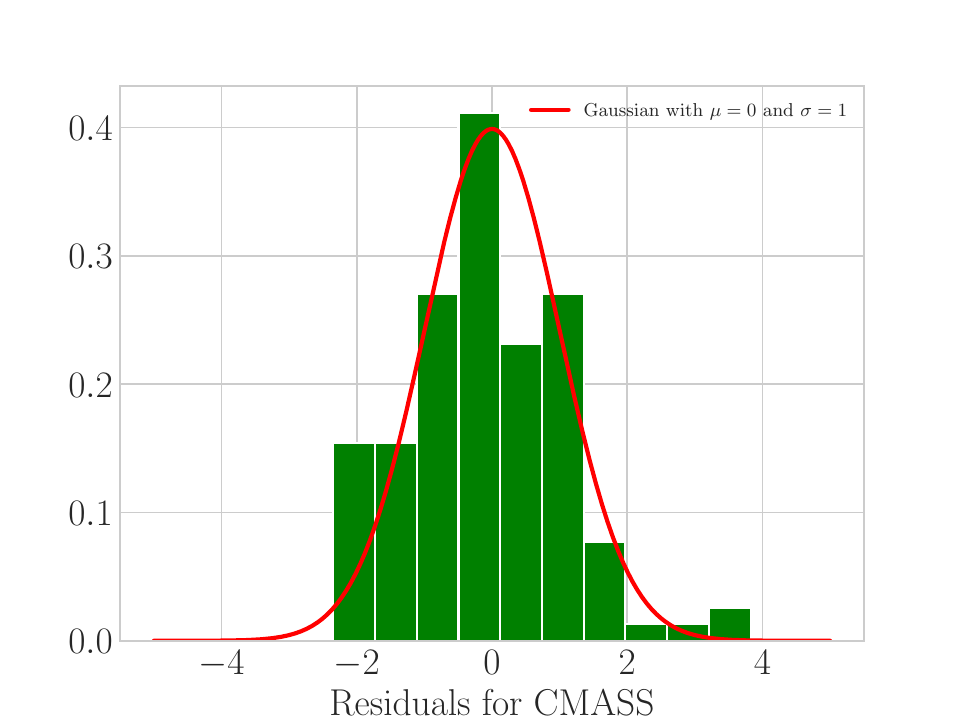}
\caption{The distribution of residuals (green histograms), for LOWZ (left panel) and CMASS (right panel), and the standard normal distribution (red curve). With the Kolmogorov-Smirnov test, we validate that both histograms are consistent with being derived from a standard normal distribution.
}
\label{fig:residuals}
\end{figure*}

In \autoref{fig:residuals}, we show the normalised residuals for LOWZ (left) and CMASS (right) after we input the mean values of the constrained parameter set \({\{\theta_\alpha\}}=\{\bs^i,\fs^i\}\). By considering a Kolmogorov-Smirnov test, we accept the null hypothesis that both histograms are consistent with being derived from a standard normal distribution since the \(p\)-values are 0.18 and 0.13 respectively at 95\% significance level, and we conclude that our model represents the data well.

\section{Conclusions}
\label{sec:conclusion}
In this paper, we have introduced a novel method to estimate the amplitude of the galaxy clustering and the growth rate of the cosmic structures with harmonic-space (tomographic) power spectra. We have tested it against synthetic data sets and simulations, and we have lastly applied it to the LOWZ and CMASS galaxy samples of the 12th data release of BOSS.

We have derived constraints on the $b\sigma_8$ and $f\sigma_8$ parameters for each redshift bin after taking into account the observational effects of the survey in a pseudo-power spectrum approach. In addition, we have constructed the data covariance with three different implementations (Gaussian, \texttt{PolSpice}, and Mock covariance matrix), all of the them yielding consistent results. On top of that, our method shows considerable independence from the fiducial theory model, and it passes successfully a series of internal consistency and systematics checks. Constraints on $b\sigma_8$ and  $f\sigma_8$ agree very well with the findings in the literature. In particular, our new method can be compared easily to the recently-proposed angular redshift fluctuations \citep{CHM2020}, and we find similar results. It could also, in principle, be extended to the follow up forecast papers as e.g.\ \citet{Fonseca_2019}.

Despite the fact that the estimated errors on our measurements are generally larger than those obtained with traditional analyses targeting the detection of RSD in Fourier or configuration space, the method described here provides complementary results and allows us to track the evolution of these fundamental cosmological quantities with time. The overwhelming wealth of data provided by forthcoming experiments is expected to improve these constraints and will hopefully shed some more light on the physics of the history of our Universe.

\section*{acknowledgments}

The authors thank Manuel Colavincenzo and Adam Hawken for collaboration in the early stages of the project, Marco Regis for enlightening discussions, and Arthur Loureiro, Isaac Tututsaus, and Amel Durakovic for feedback and support. They also thank the anonymous referee for their invaluable advice on the improvement of the presentation of the current work. KT is supported by the
European Structural and Investment Fund and the
Czech Ministry of Education, Youth and Sports (Project
CoGraDS - CZ.02.1.01/0.0/0.0/15 003/0000437). SC acknowledges support from the `Departments of Excellence 2018-2022' Grant (L.\ 232/2016) awarded by the Italian Ministry of University and Research (\textsc{mur}), and from the `Ministero degli Affari Esteri della Cooperazione Internazionale (\textsc{maeci}) -- Direzione Generale per la Promozione del Sistema Paese Progetto di Grande Rilevanza ZA18GR02.


\bibliography{bibliography}{}
\bibliographystyle{aasjournal}




\end{document}